\documentclass[12pt]{article}
\usepackage[utf8]{inputenc}
\usepackage{authblk}
\usepackage{setspace}
\usepackage[margin=1.25in]{geometry}
\usepackage{graphicx}
\usepackage{subcaption}
\usepackage{amsmath}
\usepackage{lineno}
\usepackage{lipsum}
\usepackage{wrapfig}
\usepackage[style=nejm, 
citestyle=numeric-comp,
sorting=none]{biblatex}
\begin{document}
\title{Multiplicity dependence of intra-jet properties in pp collisions at $\sqrt{s}$ = 13~TeV with ALICE}
\author[1*]{Debjani Banerjee (for the ALICE Collaboration)}
\affil[1]{Bose Institute, Kolkata, India}
\affil[*]{Address correspondence to: banerjee.debjani@cern.ch}

\onehalfspacing
\maketitle

\date{}

\begin{abstract}
Recent results in high-multiplicity pp collisions show features similar to those that are associated with the formation of a quark--gluon plasma in heavy-ion collisions. Investigating the modification of the intra-jet properties as a function of event multiplicity in pp collisions can provide deeper insight into the nature of these effects. We will present the recent measurements of multiplicity dependence of charged-particle jet properties (average charged particle multiplicity and fragmentation functions) for leading charged-particle jets. Jets are reconstructed using anti-$k_{\rm T}$ jet finding algorithm with radius parameter $R$ = 0.4 in the jet $p_{\rm T}$ range from 5 -- 110 GeV/$c$ at midrapidity in pp collisions at $\sqrt{s}$ = 13 TeV with ALICE.
\end{abstract}
\clearpage
\section{Introduction}
High-energy nuclear or hadronic collisions produce hard-scattered (large-$Q^{\rm 2}$) quarks and gluons, which fragment into a collimated spray of final state particles known as jets. Measurements of jet production in pp collisions serve as a baseline for perturbative calculations in Quantum Chromodynamics (QCD). A measurement of intra-jet properties is sensitive to details of a parton shower and hadronization processes. In addition, the measurement of jet properties as a function of event multiplicity in pp collisions is important to enrich our understanding of quark--gluon plasma-like phenomena observed in small-collision systems. In this article, we present the multiplicity dependence of intra-jet properties for charged-particle leading jets (a jet with the highest-$p_{\rm T}$ in an event), such as mean charged-particle multiplicity and fragmentation functions, measured in pp collisions at $\sqrt{s}$ = 13\,TeV with ALICE. 
\vspace*{-.2cm}
\section{Analysis details and jet observables}
The results presented here are based on the pp data collected by the ALICE detector~\cite{ALICE_det} in 2016, 2017, and 2018. Accepted events had primary interaction vertex position within $\pm 10$ cm relative to the position of the nominal interaction point along the beam axis. Minimum Bias (MB) events were selected by requiring the in-time coincidence of signals from the V0A and V0C forward scintillator arrays~\cite{V0}. High-multiplicity (HM) events were selected, when the sum of V0A and V0C amplitudes was more than 5 times greater than the mean MB signal amplitude. Charged-particle jets were reconstructed using the anti-$k_{\rm T}$ algorithm~\cite{antikT} with $p_{\rm T}$ recombination scheme in FastJet 3.2.1~\cite{FastJet} for jet resolution parameter $R$ = 0.4 from charged particles with transverse momentum ($p_{\rm T}$) $>$ 0.15 GeV/$c$ and $|\eta| <$ 0.9.  The mean charged-particle multiplicity,  $\left<N_{\rm ch}\right>$, and jet fragmentation function, $z^{\rm ch} = {p_{\rm T}^{\rm particle}}/{p_{\rm T}^{\rm jet,ch}}$ were measured for leading jets in both MB and HM events. To correct for the instrumental effects, a 2D Bayesian unfolding technique~\cite{Bayesian} implemented in the RooUnfold~\cite{RooUnfold} package was used. The underlying event (UE) contribution was estimated using the perpendicular-cone method~\cite{jetpp7TeV} and subtracted on a statistical basis after unfolding both the raw distributions and the UE contributions separately. In this analysis, the systematic uncertainties, which arise from tracking inefficiency and Monte Carlo event generator dependence, are the main contributors to the total systematic uncertainty. 
\vspace*{-.2cm}
\section{Results and discussion}
\begin{figure}[h!]
	\centering
	\begin{minipage}[b]{0.30\linewidth}
		\includegraphics[scale=0.25]{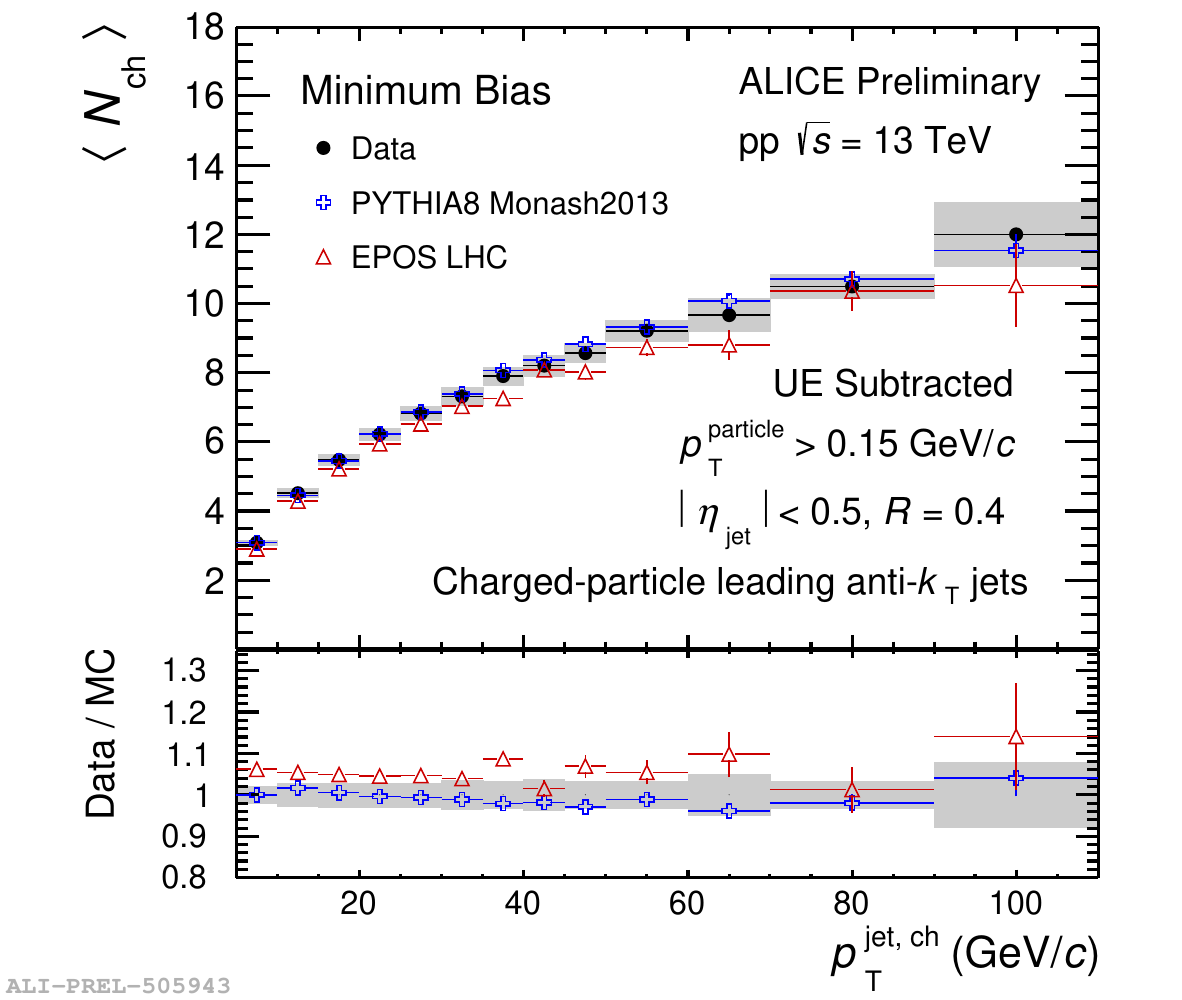}
	\end{minipage}
	\quad
	\begin{minipage}[b]{0.30\linewidth}
		\includegraphics[scale=0.27]{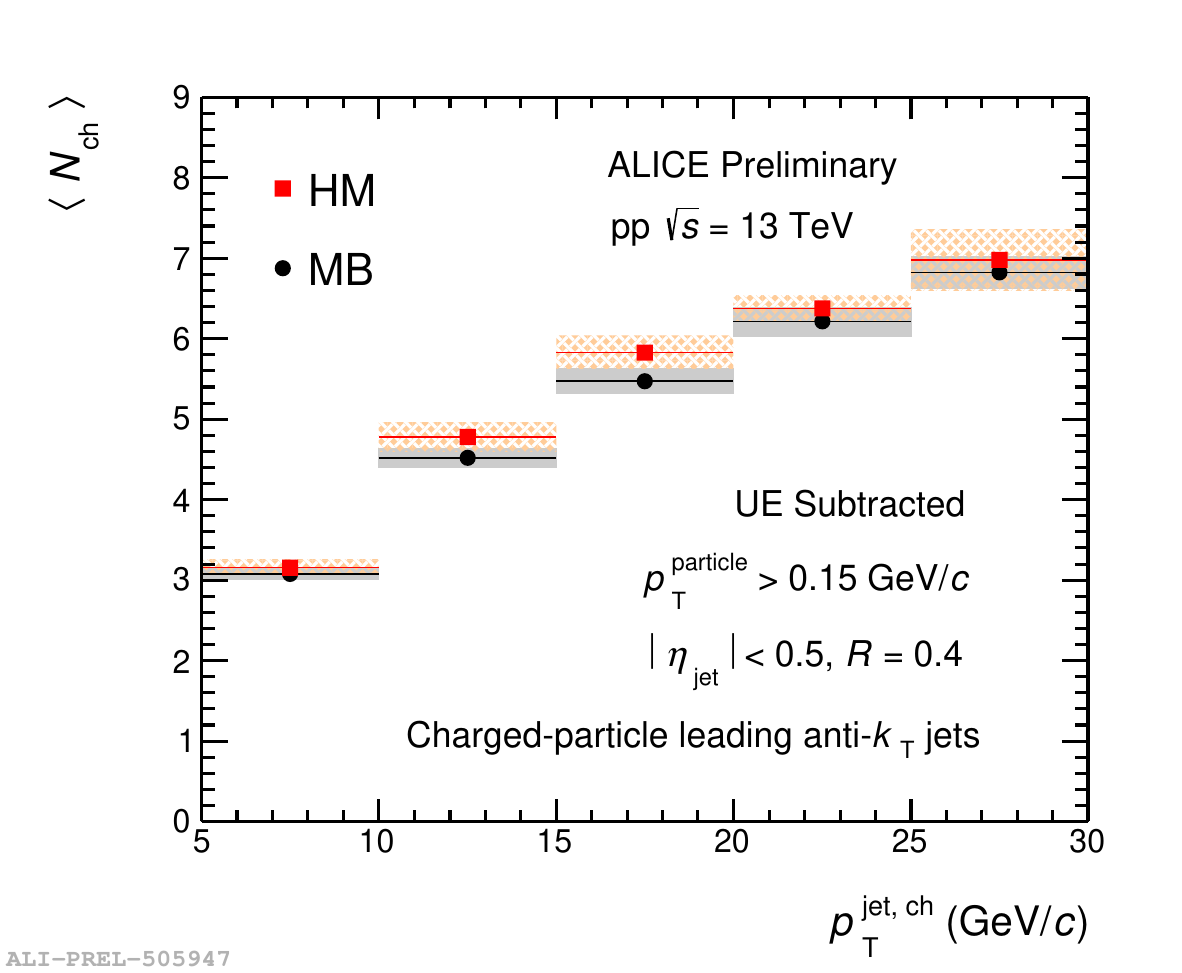}
	\end{minipage}
\quad
\begin{minipage}[b]{0.30\linewidth}
	\includegraphics[scale=0.27]{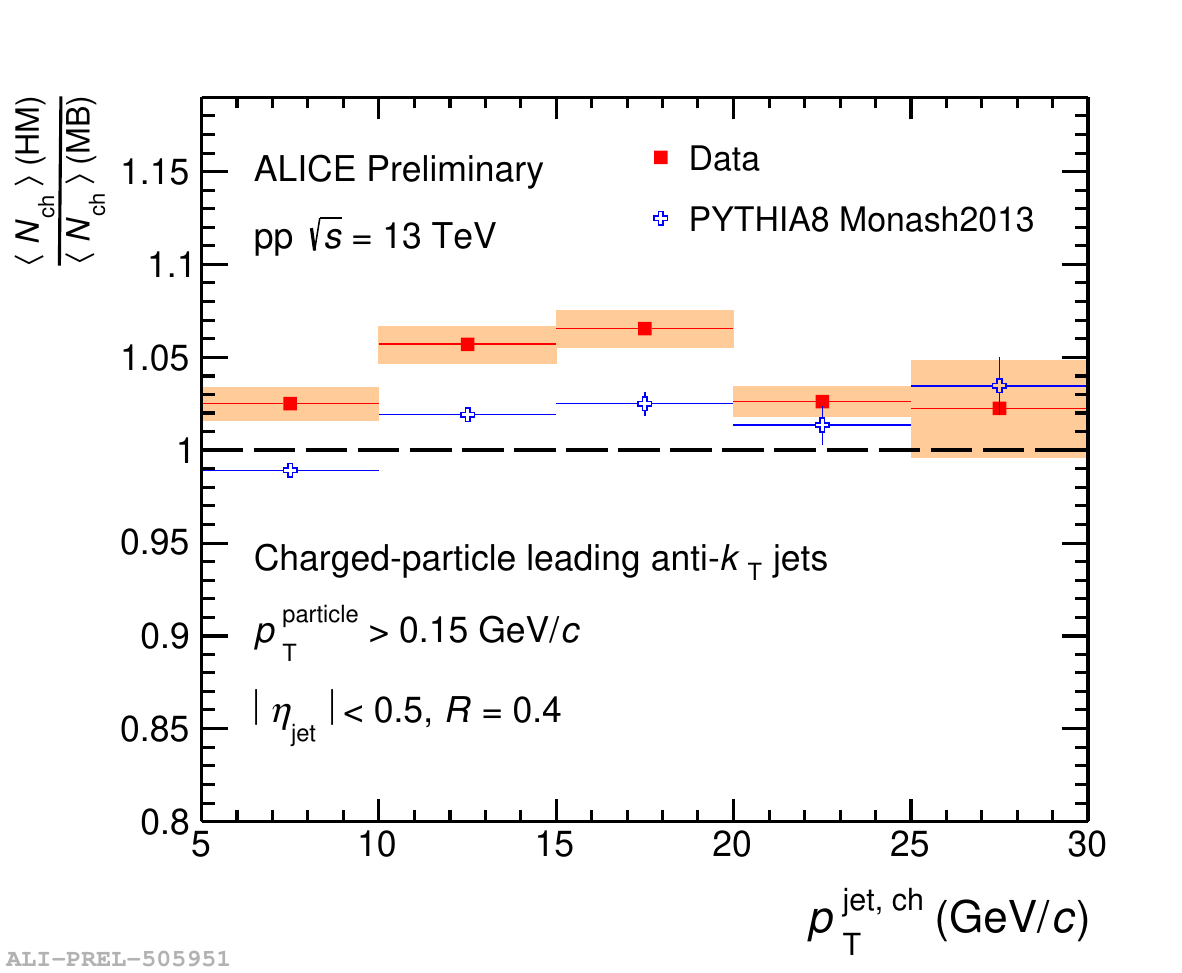}
\end{minipage}
\caption{Mean multiplicity of charged particles in a leading jet as a function of $p_{\rm T}^{\rm jet,ch}$. Left: The measured MB distribution is compared with calculations of MC event generators. Middle: Comparison of $\left<N_{\rm ch}\right>$ distributions measured in HM and MB events. Right: Ratio of $\left<N_{\rm ch}\right>$ distributions from HM and MB events.}
\label{NchHMMB}
\end{figure}
\begin{figure}[h!]
	\centering
	\begin{minipage}[b]{0.30\linewidth}
		\includegraphics[scale=0.27]{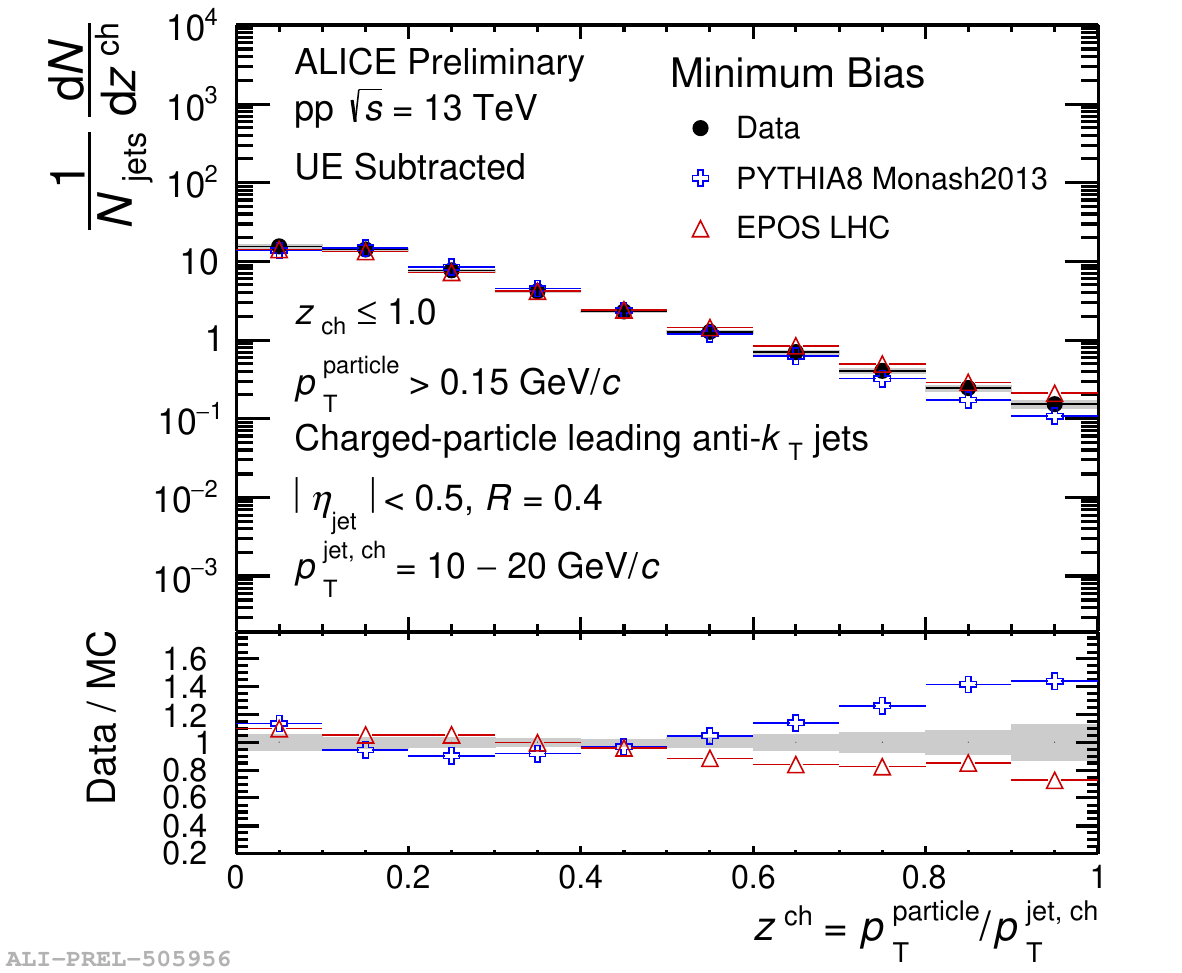}
	\end{minipage}
	\quad
	\begin{minipage}[b]{0.30\linewidth}
		\includegraphics[scale=0.27]{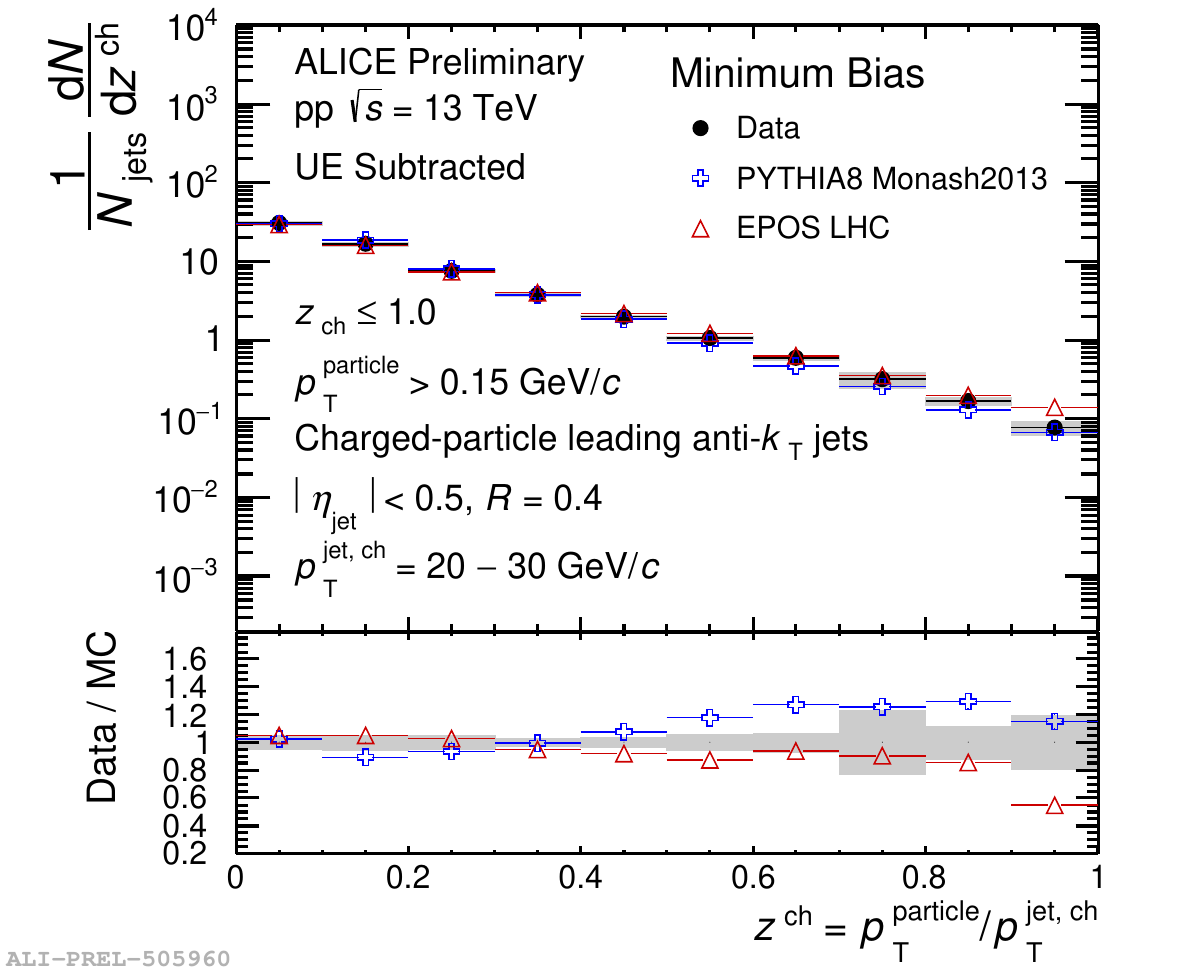}
	\end{minipage}
	\quad
	\begin{minipage}[b]{0.30\linewidth}
		\includegraphics[scale=0.27]{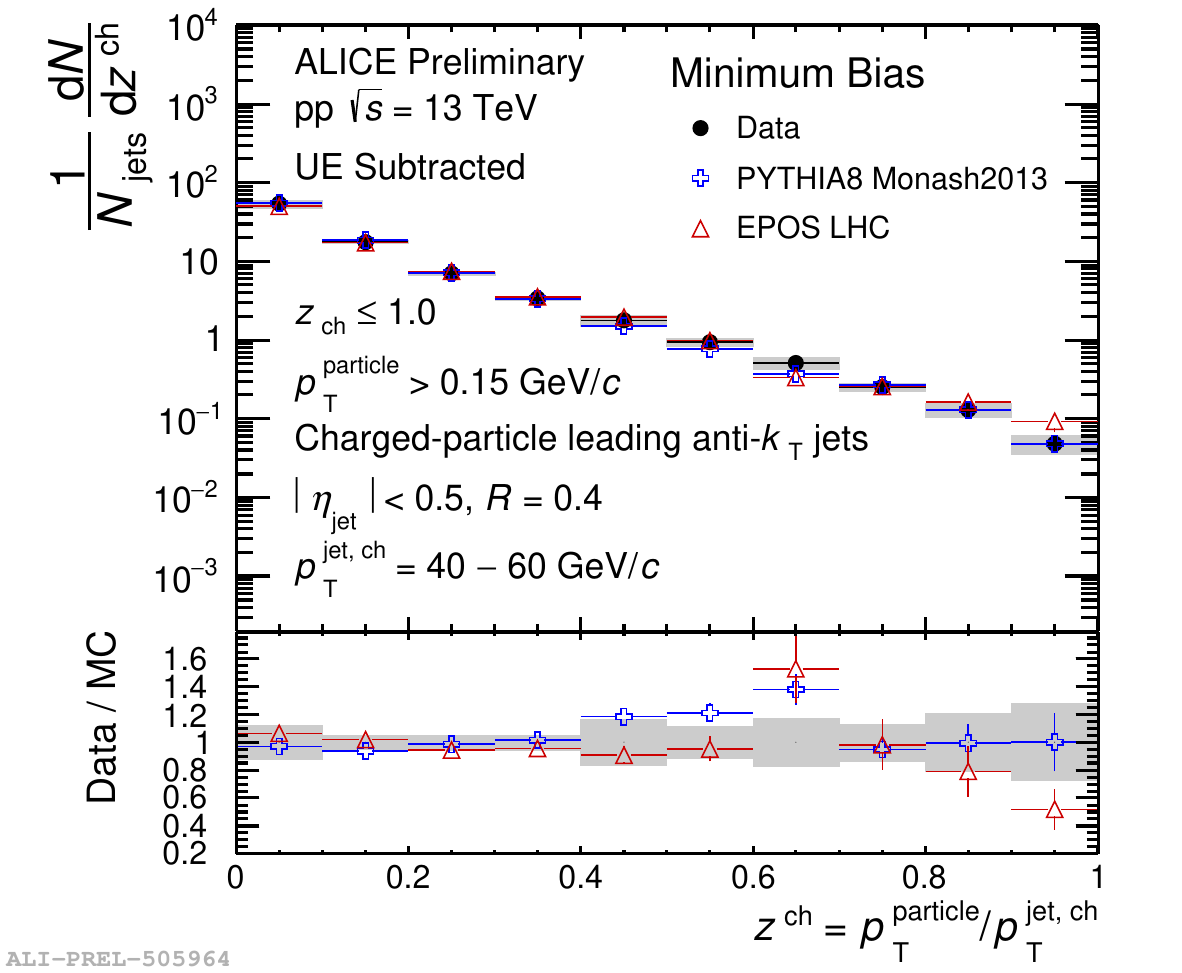}
	\end{minipage}
        \caption{Distributions of the fragmentation variable $z^{\rm ch}$ measured in three charged-particle leading-jet $p_{\rm T}$ ranges in MB events. Data are compared with calculations of MC event generators.}
	\label{zMB}
\end{figure}
Figure~\ref{NchHMMB} (left) shows $\left<N_{\rm ch}\right>$ as a function of leading-jet transverse momentum $p_{\rm T}^{\rm jet,ch}$ for MB events. The results (black markers) are compared with PYTHIA 8 Monash 2013 and EPOS LHC predictions represented by the blue and red markers, respectively. PYTHIA 8 Monash 2013 is a parton-based MC generator, where the hadronization is treated using the Lund string fragmentation model for collider physics~\cite{pythia8_monash}, whereas EPOS is based on perturbative QCD, Gribov-Regge multiple scattering, and string fragmentation~\cite{eposlhc}. The bottom panel of Fig.~\ref{NchHMMB} (left) shows ratios between the data and MC predictions. The PYTHIA 8 Monash 2013 calculation reproduces the data within systematic uncertainty while, EPOS LHC slightly underestimates the data. Figure~\ref{NchHMMB} (middle) presents $\left<N_{\rm ch}\right>$ as a function of leading jet $p_{\rm T}^{\rm jet,ch}$ for HM and MB events, represented by red and black markers for $p_{\rm T}^{\rm jet,ch} <$ 30 GeV/$c$. It is observed that $\left<N_{\rm ch}\right>$ increases with leading-jet $p_{\rm T}$ for both HM and MB events. Figure~\ref{NchHMMB} (right) shows the corresponding ratio of both distributions. It is observed that $\left<N_{\rm ch}\right>$ is slightly larger for HM compared to MB events, which is qualitatively reproduced by PYTHIA 8 Monash 2013 for all $p_{\rm T}$ except the lowest bin.
\begin{figure}[h!]
	\centering
	\begin{minipage}[b]{0.30\linewidth}
		\includegraphics[scale=0.27]{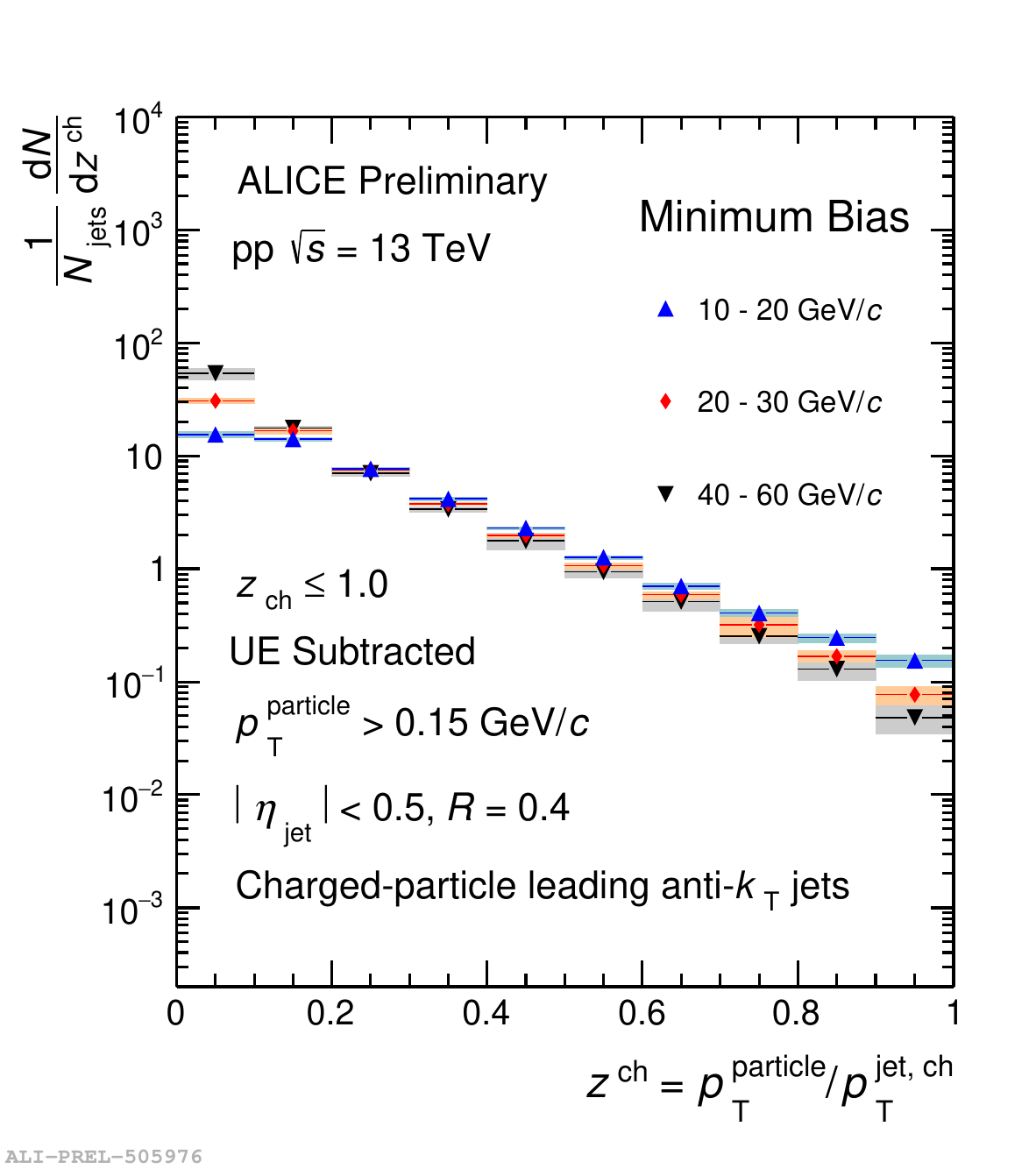}
	\end{minipage}
	\quad
	\begin{minipage}[b]{0.30\linewidth}
		\includegraphics[scale=0.27]{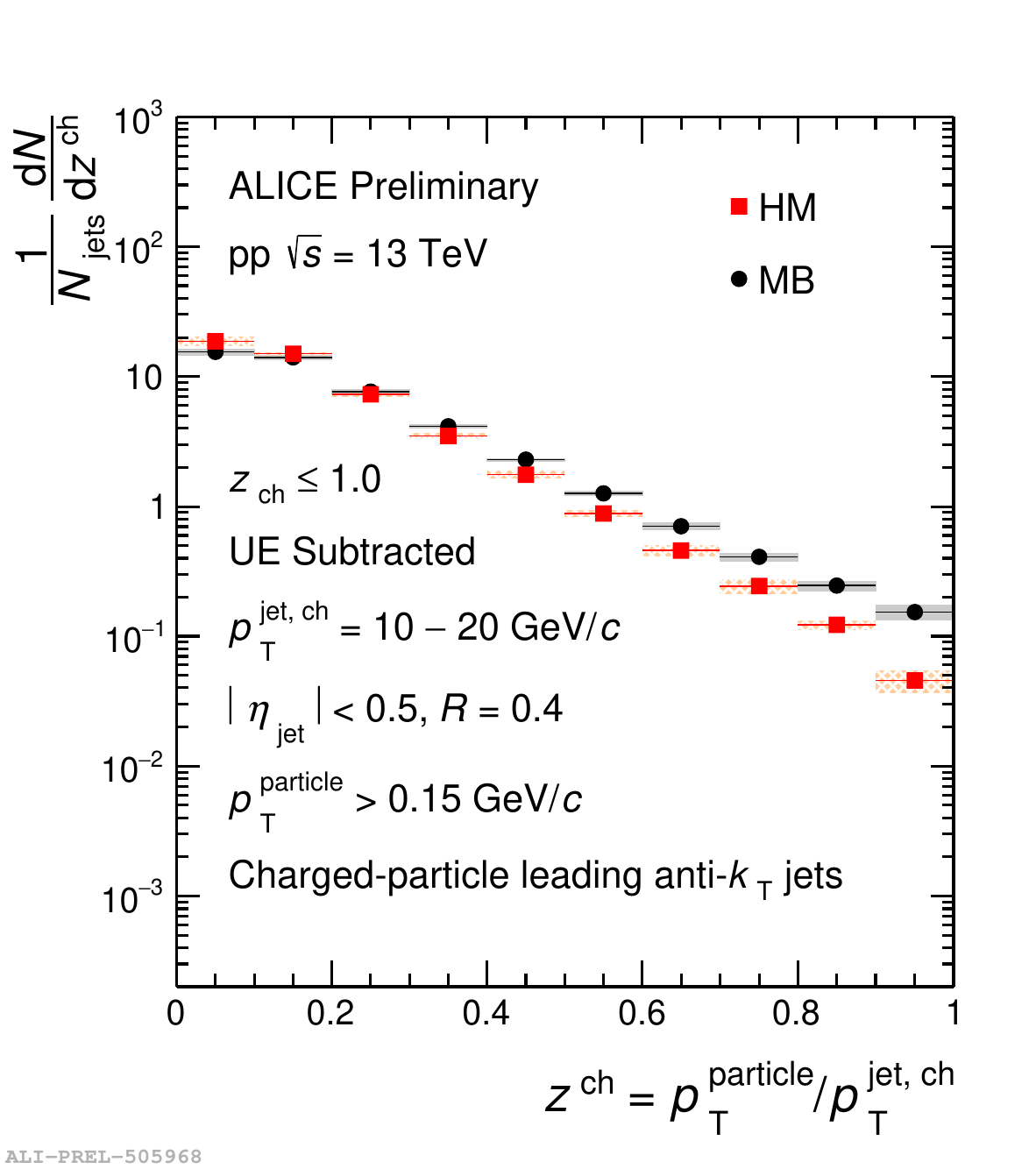}
	\end{minipage}
	\quad
	\begin{minipage}[b]{0.30\linewidth}
		\includegraphics[scale=0.27]{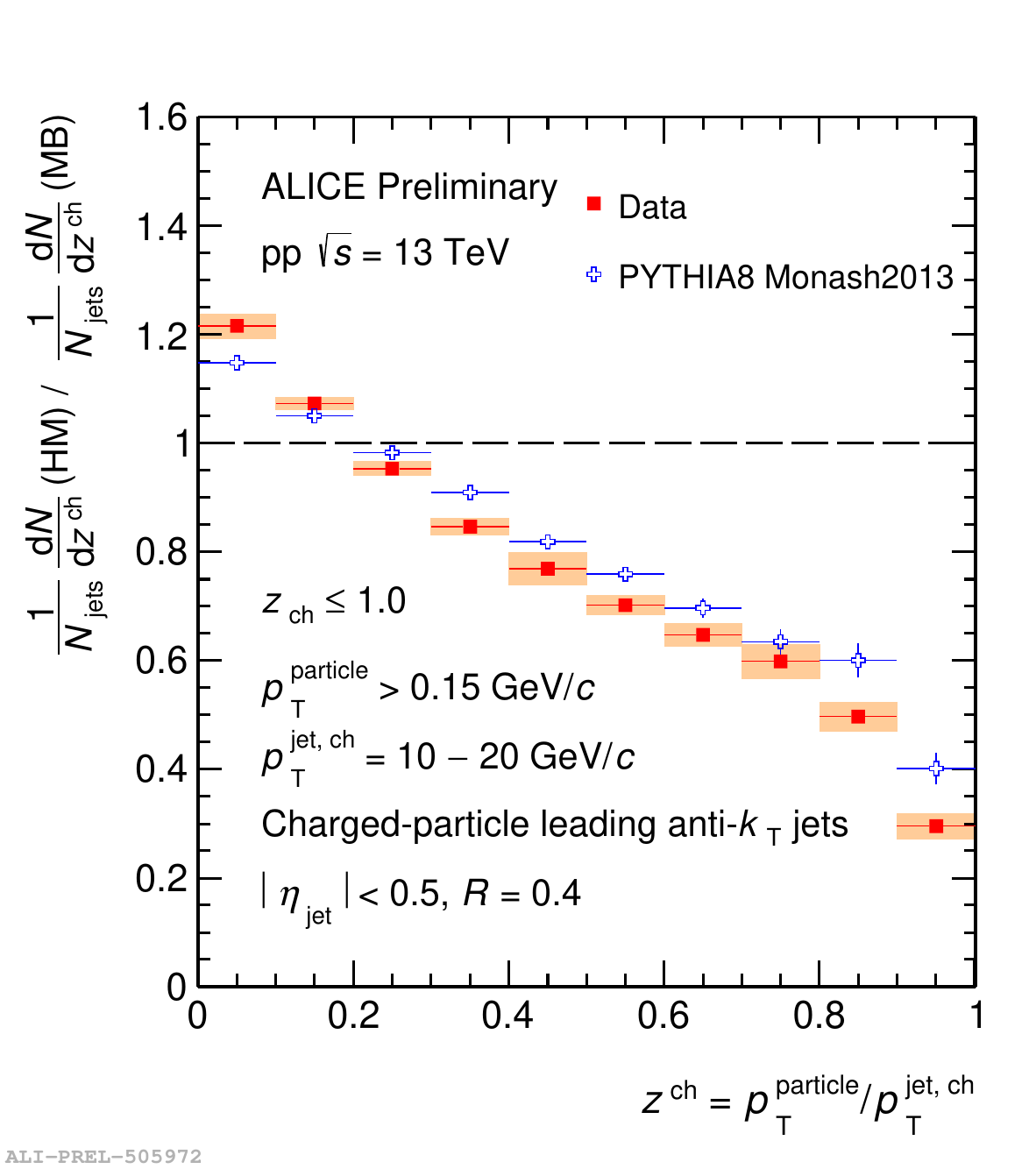}
	\end{minipage}
	\caption{Left: Measured $z^{\rm ch}$ distributions in three charged-particle leading-jet $p_{\rm T}$ ranges in MB events. Middle: Comparison of the MB and HM $z^{\rm ch}$ distributions for leading jets with $p_{\rm T}$ = 10 -- 20 GeV/c. Right: Ratio of both distributions.}
	\label{zallMBHM}
      \end{figure}
      
Figure~\ref{zMB} shows $z^{\rm ch}$ distributions for leading jets with $p_{\rm T}^{\rm jet,ch}$ = 10 -- 20 GeV/$c$ (left), 20 -- 30 GeV/$c$ (middle), and 40 -- 60 GeV/$c$ (right), and compares the measurements with EPOS LHC  (red markers) and PYTHIA 8 Monash 2013 (blue markers) predictions. The bottom panels in Fig.~\ref{zMB} show the corresponding ratios.

It is observed that both models reproduce the data at low $z^{\rm ch}$ ($<$ 0.5), whereas for high $z^{\rm ch}$ ($>$ 0.5) and lower jet $p_{\rm T}$ range, EPOS LHC describes the data better than PYTHIA 8 Monash 2013. However, for high $z^{\rm ch}$ ($>$ 0.5) and higher jet $p_{\rm T}$ (40 -- 60~GeV/$c$), both models are compatible with the data within systematic uncertainties. Figure~\ref{zallMBHM} (left) shows the MB $z^{\rm ch}$ distributions corresponding to the three jet-$p_{\rm T}$ ranges. It is observed that except at the highest and lowest $z^{\rm ch}$ value, the distributions are consistent within systematic uncertainties, indicating scaling of the fragmentation function independent of jet $p_{\rm T}$. Figure~\ref{zallMBHM} (middle) presents $z^{\rm ch}$ distributions for HM and MB events, marked with red and black markers, respectively and Fig.~\ref{zallMBHM} (right) shows their ratio. An enhancement (suppression) is observed for low (high) $z^{\rm ch}$ in HM events compared to MB, however, PYTHIA 8 Monash 2013 qualitatively follows the similar trend.

\vspace*{-.2cm}
\section{Summary}
Charged-particle intra-jet properties were measured for minimum-bias and high-multiplicity events in pp collisions at $\sqrt{s}$ = 13\,TeV with ALICE. A slight enhancement in $\left<N_{\rm ch}\right>$ is observed in HM events compared to that in MB events. Modification of $z^{\rm ch}$ distributions is observed in HM events compared to that in MB events and this feature is qualitatively reproduced by PYTHIA 8 with the Monash 2013 tune, indicating softening of jet fragmentation in HM compared to MB.
\printbibliography

\end{document}